\documentclass[aps,prd,twocolumn,preprintnumbers,groupedaddress,superscriptaddress,floatfix,tightenlines,reprint,nofootinbib,longbibliography]{revtex4-1}
\usepackage{mathrsfs,natbib}
\usepackage{verbatim}
\usepackage{bm}
\usepackage{enumerate}
\usepackage{graphicx,amssymb,bbm,amsbsy,amsfonts,amssymb,amsmath}
\usepackage{hyperref}

\def\tr{\text{tr}\,}
\def\eq{\,{=}\,}

\def\eq{\,{=}\,}

\hypersetup{
    colorlinks=true,       
    linkcolor=red,          
    citecolor=blue,        
    filecolor=magenta,      
    urlcolor=blue           
}

\begin{document}

\title{Anyonic particle-vortex statistics and the nature of dense quark matter}

\author{Aleksey Cherman}
\email{aleksey.cherman.physics@gmail.com}
\affiliation{Institute for Nuclear Theory, University of Washington, Seattle, WA 98195 USA}
\altaffiliation{Permanent address:  School of Physics and Astronomy, University of Minnesota, Minneapolis, MN 554555 USA}
\author{Srimoyee Sen}
\email{srimoyee08@gmail.com}
\affiliation{Institute for Nuclear Theory, University of Washington, Seattle, WA 98195 USA}
\author{Laurence G. Yaffe}
\email{yaffe@phys.washington.edu}
\affiliation{Department of Physics, University of Washington, Seattle, WA 98195 USA}

\preprint{INT-PUB-18-043}

\begin{abstract}
We show that $\mathbb{Z}_3$-valued particle-vortex braiding phases
are present in high density quark matter. Certain mesonic and baryonic
excitations, in the presence of a superfluid vortex, have orbital angular
momentum quantized in units of $\hbar/3$. Such non-local topological
features can distinguish phases whose realizations of global symmetries,
as probed by local order parameters, are identical. If $\mathbb{Z}_3$
braiding phases and angular momentum fractionalization are absent in
lower density hadronic matter, as is widely expected, then the quark
matter and hadronic matter regimes of dense QCD must be separated by at
least one phase transition.
\end{abstract}

\maketitle
\flushbottom

\section {Introduction}
The behavior of QCD as a function of baryon density, at vanishing temperature,
is of fundamental importance to nuclear physics and astrophysics
\cite{doi:10.1146/annurev.ns.21.120171.000521,
    Fukushima:2010bq,
    Ozel:2016oaf,
    Baym:2017whm}.
At low density, quarks and gluons have strong interactions and are bound
into colorless hadrons, producing hadronic nuclear matter.
At asymptotically high densities, one instead finds weakly-coupled quark matter
\cite{Alford:2007xm}.
Can the hadronic and quark matter regimes be smoothly connected,
or are they necessarily separated by a phase transition?

We consider this question in the simplified,
more symmetric setting of three-color, three-flavor QCD with
degenerate quark masses and $SU(3)_V$ flavor symmetry.%
\footnote
    {%
    When electromagnetic and weak interactions are included,
    and quarks have their physical masses, various
    phase transitions associated with, e.g., kaon condensation 
    occur at high, but non-asymptotic, density \cite{Bedaque:2001je,Kaplan:2001qk,Kryjevski:2004cw,Alford:2007xm}.
    Focusing on the flavor symmetric limit avoids these complications.
    }
The increase in symmetry gives some hope that questions of principle
can be addressed in a sharper fashion.
In this flavor-symmetric setting, there is a well-known
conjecture of \emph{quark-hadron continuity} due to Sch\"afer and Wilczek
\cite{Schafer:1998ef}.
This conjecture is supported by a comparison of the expected pattern of
low energy excitations and the realizations of conventional global symmetries
at both high and low densities.

At asymptotically high densities,
`color superconductivity' leads to a diquark `condensate'
$\langle qq \rangle \neq 0$,
which in turn induces non-zero gauge-invariant condensates
with the schematic forms $\langle (qq)^3 \rangle$ and
$\langle \bar{q}\bar{q}qq \rangle$ \cite{Alford:1998mk}.
The $\langle (qq)^3 \rangle$ condensate signals spontaneous
breaking of the $U(1)_B$ baryon number symmetry down to $\mathbb Z_2$,
while the $\langle \bar{q}\bar{q}qq \rangle$ condensate signals,
in the limit of massless quarks,
spontaneous chiral symmetry breaking of the form
$SU(3)_L \times SU(3)_R \to SU(3)_{V}$.
The spontaneous breaking of $U(1)_B$ indicates that
high density quark matter is a superfluid.
At low densities, one expects an identical chiral symmetry breaking pattern
in the massless limit, while $U(1)_B$ symmetry breaking is believed to arise
(in the flavor symmetric theory) from condensation of pairs of $\Lambda$
hyperons with flavor content $uds$.
The matching symmetry realization, along with apparently compatible
patterns of
low energy excitations, make it plausible that
the quark and hadronic phases of QCD are smoothly connected,
at least in the flavor symmetric limit \cite{Schafer:1998ef}.

However, there is no guarantee that distinct phases can always be
distinguished by this Landau-Ginzburg style analysis based on
local order parameters.
Some transitions separating distinct phases can only be diagnosed
by changes in behavior of topological observables such as the
ground state degeneracy on large topologically non-trivial manifolds,
or suitable non-local order parameters from which one infers,
for example, particle-vortex braiding statistics
\cite{doi:10.1142/S0217979290000139,
    PhysRevB.40.7387,
    PhysRevB.41.9377}.

Given this motivation, we examine topological ground state
degeneracies and quark-vortex braiding statistics in asymptotically
high-density quark matter,
and compare results with the expected properties
of hadronic nuclear matter.
In high density quark matter, we find that quarks acquire
non-trivial $\mathbb{Z}_3$ Aharonov-Bohm
phases, arising from color holonomies,
when encircling a superfluid vortex with minimal circulation.
In terms of dressed gauge-invariant quasiparticle excitations, this means that
certain mesonic and baryonic excitations
have orbital angular momentum quantized in units of $\hbar/3$
in the presence of a minimal superfluid vortex.  These results can also be interpreted in terms of anyonic particle-vortex braiding statistics.

These topological features contrast sharply with the expected
properties of superfluid hadronic matter,
in which one expects conventional quantization of quasiparticle orbital
angular momentum in units of $\hbar$.
If the standard picture of the low density hadronic regime is correct,
then the hadronic and quark matter
regimes must be separated by at least one phase transition.

\section {\boldmath $U(1)$ superconductors}
To set the stage for our QCD discussion we first review related
aspects of BCS superconductors at zero temperature.
(See, for example, Ref.~\cite{Hansson:2004wca} for more detail.)
Such systems can be modeled by an Abelian Higgs model,
\begin{align}
    \mathcal{L} =
    |D_{\mu} \phi|^2
    + m^2 |\phi|^2
    + \tfrac{1}{2}\lambda  |\phi|^4
    - \frac{1}{4 e^2} F^2_{\mu \nu} \,.
\label{eq:AbelianHiggs}
\end{align}
The complex scalar field $\phi$ is assumed to have charge $q$
under the $U(1)$ gauge symmetry,
so $D_{\mu}\phi \equiv (\partial_{\mu} -  i q A_{\mu}) \phi$,
with $q \eq 2$ for real electron superconductors.
Using sloppy perturbative language,
$\phi$ gets a non-zero vacuum expectation value
in the superconducting regime of large negative $m^2$,
`breaking' the $U(1)$ gauge symmetry.
But local
gauge symmetries never truly break spontaneously \cite{Elitzur:1975im}.
Realizations of all conventional global symmetries remain unchanged
as $m^2$ varies, and there are no gauge-invariant local operators
which can serve as
order parameters for superconductivity.
Distinguishing the superconducting and normal phases requires
looking beyond the conventional Landau-Ginzburg paradigm based
on local order parameters.

The superconducting phase has $\mathbb{Z}_2$ topological order,
arising from Higgsing of the $U(1)$ gauge symmetry down to
$\mathbb{Z}_2$ \cite{Hansson:2004wca}.
This provides a sharp distinction from the normal phase.
In this context, topological order
\cite{doi:10.1142/S0217979290000139,PhysRevB.40.7387,PhysRevB.41.9377} has
two related consequences:
a ground state degeneracy which depends on the topology of space,
and non-trivial Aharonov-Bohm phases for transport of charged particles
around magnetic vortex lines.

To examine the ground state degeneracy, one may compactify a single
spatial direction with periodic boundary conditions (for all fields),
and consider the theory on $\mathbb{R}^{1,2} \times S^1$.
Ground state degeneracy arises from multiple minima of the
holonomy effective potential $V_{\rm eff} (\Omega)$, where
$\Omega \equiv e^{i \oint A}$ is the
Wilson loop (or spatial holonomy) wrapping the compactified dimension.
In the Higgs phase, the covariant gradient term of the Lagrangian 
(\ref{eq:AbelianHiggs}) induces a (Meissner) mass for the photon.
When the spatial holonomy $\Omega \ne 1$, this term also gives
a tree level contribution to the holonomy potential,
\begin{align}
    V_{\rm eff} (\Omega)
    =
    \min_{k \in \mathbb{Z}} \frac{|v|^2}{L^2}\left(2\pi k -q a\right)^2 + \cdots \,,
\label{eq:abelianHiggs}
\end{align}
where $\Omega \equiv e^{i a}$, 
the ellipsis denotes higher order contributions
(plus terms independent of $\Omega$),
$v^2 \equiv -m^2/\lambda$,
and $k$ is the winding number of the phase of the condensate
around the $S^1$.
Viewed as a function of $a = -i \ln \Omega$,
gauge invariance requires the  potential to be $2\pi$ periodic,
but when $q > 1$ it actually has a finer periodicity of $2\pi/q$.
For $q \eq 2$ there are two degenerate minima within the 
fundamental domain $[0,2\pi)$,
namely $a \eq 0$ and $a\eq\pi$,
associated with $k \eq 0$ and $k\eq1$, respectively.
So $\langle\Omega\rangle = \pm 1$ and the ground state degeneracy is 2
on $\mathbb{R}^{1,2} \times S^1$.
(On $\mathbb R \times T^3$, with charge $q$, the degeneracy is $q^3$.)

While one does not directly measure this ground state degeneracy
experimentally,
the $\Omega \eq -1$ minimum is related to the 
braiding statistics between charged particles and magnetic vortices
\cite{PhysRevLett.56.792}.
The field configuration describing a superconducting vortex running along some
straight path is, in cylindrical coordinates, given by
\begin{align}
    \phi(\theta,r) = f(r) \,  e^{i k \theta}, \quad
    A_{\theta} = a \, {h(r)}/{r} \,,
\end{align}
where the radial functions $f(r)$ and $h(r)$ run monotonically
from 0 at $r\eq 0$ to 1 at $r\eq\infty$, and
$k \in \mathbb{Z}$.
For the vortex to have finite energy per unit length,
the covariant derivative of $\phi$ must vanish at large $r$,
implying that $a = k/2$.
So a minimal vortex carrying $k\eq 1$ units of magnetic flux has an
azimuthal gauge field $A_\theta \sim 1/(2r)$ at large $r$,  
implying that the holonomy around a large circle surrounding the vortex
is $-1$.
More generally, if $\Omega[C]$ denotes the holonomy (Wilson loop)
around some closed loop $C$, then
\begin{align}
    \langle \Omega[C]\rangle_{V[P]} = e^{i \pi \ell(C,P)}\, ,
\label{eq:linking_BCS}
\end{align}
where $\langle \cdots \rangle_{V[P]}$ denotes an expectation value
in the presence of a vortex $V$ running along some closed path $P$, 
and $\ell(C,P)$ is the intersection (linking) number of paths $C$ and $P$,
provided these paths are well separated.%
\footnote
    {%
    When fluctuations are taken into account, $|\langle \Omega \rangle| $ has perimeter law decay, and
    the left-hand side of expression (\ref{eq:linking_BCS}) should
    really be the phase of the Wilson loop, or equivalently
    the ratio
    $
    \langle \Omega[C]\rangle_{V[P]}/ \langle \Omega[C]\rangle
    $, 
    where the denominator is the Wilson loop expectation value
    in the absence of the vortex.
    }
This shows that particles of charge $q \eq 1$ have an Aharonov-Bohm
phase of $-1$ when going around a vortex, demonstrating
that particles and vortices have $\mathbb{Z}_2$ braiding statistics.
To reiterate, the holonomy
$\Omega[C]$ may be far from unity in the presence
of vortices, or with non-trivial topology, 
even though electric and magnetic fields are completely screened
in a superconductor \cite{Reznik:1993cw,Hansson:2004wca}.

\section {High density QCD}
We now generalize this analysis to the case of QCD with
$SU(3)$ flavor symmetry.
Purely for simplicity of presentation,
we assume that the quarks are massive, $m_q > 0$,
so that the theory has a non-vanishing mass gap.
(We comment on the chiral limit below.)
We consider the limit of asymptotically large
quark density, and vanishing temperature, so that
the quark chemical potential $\mu$
(equal to 1/3 times the baryon chemical potential $\mu_B$)
is large compared to the strong scale $\Lambda$
or the light quark mass $m_q$.
This regime is weakly coupled at all length scales,
thanks to Cooper pairing of quarks at the Fermi surface
and the resulting color Meissner effect which suppresses
long wavelength gauge field fluctuations.
In gauge-variant language, the large $\mu$, low temperature regime features
a diquark condensate of the `color-flavor-locked' (CFL) form,
\begin{align}
    \langle q^i_{a} \, C	\, q^j_{b} \rangle
    = K \, g(\mu)^{-1} \, \mu^2 \Delta \, \epsilon^{ijk} \, \epsilon_{abk}
    \equiv \Phi^{ij}_{ab} \,,
\label{eq:cond1}
\end{align}
where $K$ is a pure numerical factor, indices $i, j$ denote color,
$a, b$ are flavor indices, and 
$\Delta \sim \mu \, g(\mu)^{-5} e^{-3\pi^2/(g(\mu)\sqrt{2})}$
 is the pairing gap \cite{Schafer:1999fe,Alford:2007xm}.
Equivalently,
at long distances there are three emergent anti-fundamental 
Higgs fields, which may be viewed as forming a single $3\times3$ matrix
valued scalar field,
\begin{align}
    (\phi)^l_m
    \equiv
    (4K)^{-1} \, g(\mu) \, \mu^{-2} \, \epsilon^{ijl} \, \epsilon_{abm}
    \, \Phi^{ij}_{ab} \,.
\label{eq:cond2}
\end{align}
In the unitary gauge of Eq.~\eqref{eq:cond1},
$\phi = \Delta \cdot \mathbbm{1}$.

\begin{figure*}[th]
\centering
\includegraphics[width=.6\textwidth]{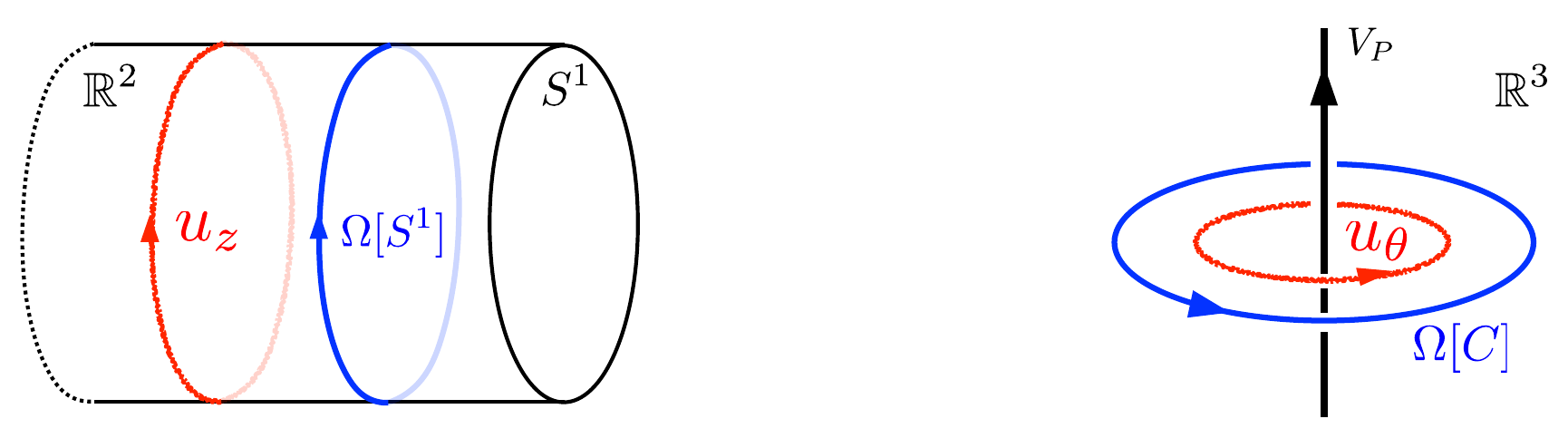}
\caption
    {%
    Left:
    $SU(3)$ holonomy $\Omega$ wrapping the compactified circle
    on spatial manifold $\mathbb{R}^2 \times S^1$ in dense quark matter.
    Non-trivial $\mathbb Z_3$-valued holonomies
    are associated with non-zero superfluid flow around the $S^1$.
    Right:
    the related 
    situation of a Wilson loop on curve $C$
    encircling a minimal superfluid vortex on path $P$,
    with linking number $\ell(C,P) \eq 1$. 
    }
\label{fig:holonomy_linking}
\end{figure*}

The determinant of $\phi$ is the physical, gauge invariant
order parameter which was earlier written schematically
as $\langle (qq)^3 \rangle$.
Under the $U(1)_B$ baryon number symmetry, $\det \phi$ has charge $2$.
Since $\langle \det \phi \rangle \neq 0$,  $U(1)_B$ symmetry
is spontaneously broken down to $\mathbb{Z}_2$
and dense quark matter is a superfluid.
Fluctuations in the condensate phase $\varphi \equiv -i \log \det \phi$
represent $U(1)_B$ Nambu-Goldstone bosons (NGBs), with associated
the low energy effective action
\begin{align}
    S_{U(1)_B} = \int d^4x \>
    \tfrac{1}{2} f^2
    \left[
	(\partial_{t} \varphi)^2
	+ v_{\rm s}^2 (\nabla \varphi)^2
    \right] \,.
\label{eq:naiveEFT}
\end{align}
Here $f^2 \sim 6 \mu^2/\pi^2$ and $v_{\rm s}^2 \sim 1/3$ \cite{Son:1999cm}.

The action~\eqref{eq:naiveEFT} is conventionally believed to be a valid
long-distance effective action for dense QCD when $m_q \,{>}\, 0$.
In the massless limit,  $m_q \,{\to}\, 0$,
there are additional Nambu-Goldstone bosons
(pions, kaons, etc.) due to the spontaneous breaking of chiral symmetry
\cite{Casalbuoni:1999wu}.
However, at the level of two derivative terms in the effective action
there is no coupling between phase fluctuations of the
neutral superfluid condensate and these chiral symmetry NGBs.
Consequently, such chiral symmetry NGBs play no role in any of the
phenomena we discuss below, and may simply be ignored.

\section {Color holonomies}\label{sec:holonomies}
We would like to examine possible equilibrium states 
on topologically non-trivial spatial manifolds, and understand
the related issue of particle-vortex braiding statistics.
To this end, we follow the same procedure used above.
We compactify one spatial dimension on a circle of circumference
$L$, with periodic boundary conditions for all fields
and $L$ larger than all intrinsic length scales.%
\footnote
    {%
    On the resulting $\mathbb{R}^2\times S^1$ spatial manifold,
    spontaneous breaking of continuous symmetries
    remains possible at zero temperature.
    }
Fluctuations in the condensate (\ref{eq:cond2}),
as well as the color Meissner effect,
may be described by an effective Lagrangian,
\begin{align}
    \mathcal{L}_{\rm \phi}
    =  \kappa \, \tr \big[
	    (D_{t} \phi)^\dagger (D_{t} \phi)
	    + v_{\rm s}^2 \, (D_{i} \phi)^\dagger (D_{i} \phi)
	\big]
    + V(\phi) \,,
\label{gin}
\end{align}
where
$\kappa = O(\mu^2/\Delta^2)$,
$D_{\mu} = \partial_{\mu} + i A_{\mu} $ is the color covariant
derivative for an antifundamental,
and the scalar potential $V(\phi)$ is minimized when
$\phi/\Delta \in U(3)$ \cite{Casalbuoni:1999wu}.

Let $\Omega$ denote the $SU(3)$ color holonomy around the $S^1$.
The gradient term in the Lagrangian (\ref{gin})
generates a tree-level contribution to 
the holonomy effective potential $V_{\rm eff}(\Omega)$.%
\footnote
    {%
    Ref.~\cite{Cherman:2017tey} examined one-loop contributions
    to the holonomy effective potential in dense QCD with flavor-twisted
    boundary conditions, 
    but did not include this dominant tree-level contribution.
    Correcting this oversight eliminates the infinite sequence
    of alternating transitions discussed in Ref.~\cite{Cherman:2017tey}.
    }
One finds
\begin{align}
    V_{\rm eff} (\Omega)
    = \frac{\kappa \Delta^2 v_s^2}{L^2}
    \min_{k \in \mathbb{Z}^3}
    \sum_{i=1}^{3} \> (2\pi k_i +\theta_i)^2  
    + \cdots \,,
\label{eq:QCDV}
\end{align}
where $\theta = (\theta_1, \theta_2, \theta_3)$
are the phases of the eigenvalues of $\Omega$,
with $\theta_3 \equiv -\theta_1{-}\theta_2$,
and $k \equiv (k_1,k_2,k_3)$ are the winding numbers of the
eigenvalues of $\phi$ around the $S^1$.

Within the triangular fundamental domain given by
$ -2\theta_1 \le \theta_2 \le \theta_1$,
$ \theta_2\le 2\pi - 2 \theta_1$,
and $ 0 \le \theta_1\le 4\pi/3 $,
the holonomy potential has
one global minimum and two degenerate local minima.
The global minimum lies at $\theta= (0,0,0)$ and is
associated with $k = (0,0,0)$.
The two local minima are $\theta = (2\pi/3,2\pi/3,-4\pi/3)$
and $\theta = (4\pi/3,-2\pi/3,-2\pi/3)$, and are associated with
$k = (0,0,1)$ and $(-1,0,0)$, respectively.
In other words, the global minimum is $\Omega = \mathbbm 1$,
while the local minima lie at $\Omega = e^{\pm 2\pi i/3} \, \mathbbm 1$.
The energy density at the local minima exceeds that at the global
minimum by the amount
\begin{equation}
    \Delta V_{\rm eff} = \frac{4\pi^2 \kappa\Delta^2 v_s^2 }{3L^2} \,.
\end{equation}

Unlike the case of a BCS superconductor, high density QCD on
$\mathbb R^3 \times S^1$ has a unique ground state, so the color superconducting state does not have conventional topological order.  
But local minima are present in which the holonomy is a
non-trivial cube root of unity, with energy density (relative to
the ground state) vanishing as $L \to \infty$, but only as
an inverse power of $L$.
This differs from generic metastable ``false vacua'' whose energy
is linear in the spatial volume,
and is also unlike the exponentially small (in $L$) energy differences
typically present in gapped systems with degenerate infinite volume
ground states.

To elucidate the physical interpretation of these non-trivial local minima,
consider the superfluid flow velocity $u_\mu$, given by
the gradient of the phase of the gauge invariant condensate $\det\phi$
divided by twice the baryon chemical potential $\mu_B$.
Equivalently,
$
    u_\mu = (2\mu_B)^{-1} \, \tr \phi^{-1} D_{\mu} \phi
$.
Evaluated in the homogeneous states described by the holonomy effective potential,
one finds a superfluid velocity along the compactified direction given by
\begin{align}
    u_z = \frac{\pi}{\mu_B L} \, (k_1+k_2+k_3) \,.
\label{eq:supercurrent}
\end{align}
The $\Omega \eq \mathbbm 1$ global minimum has vanishing superfluid velocity,
but the non-trivial local minima with $\Omega = e^{\pm 2\pi i/3} \, \mathbbm{1}$
have $\sum_i k_i = \pm 1$, and hence have supercurrents flowing around the
compact direction with the minimal non-zero quantized circulation.
The same quantized circulation appears around superfluid vortices,
and one may regard the circle-compactified theory as mimicking an annular
region surrounding a superfluid vortex,
as illustrated in Fig.~\ref{fig:holonomy_linking}.

\section {Vortices in dense quark matter}
Minimal circulation vortex configurations in high density quark matter were first
examined in Ref.~\cite{Balachandran:2005ev}.%
\footnote
    {%
    These vortices have variously been termed `semilocal' and
    `non-Abelian' \cite{Eto:2009kg,Eto:2013hoa,Gorsky:2011hd,Eto:2011mk}.
    The former terminology emphasizes that they are magnetic vortices
    as far as the $SU(3)$ gauge group is concerned but at the same time are
    superfluid vortices with logarithmically divergent energy per unit length.
    The latter terminology refers to the gauge-dependent
    notion of color-magnetic flux.
    }
In the CFL regime, a suitable field ansatz for describing a minimal
vortex on a path $P$ running along the $z$ axis is given by
\cite{Balachandran:2005ev}:
\begin{subequations}%
\label{eq:CFL_vortex}%
\begin{align}
    \frac{\phi}{\Delta} &= \textrm{diag}[e^{i\theta} f(r),\, g(r),\,g(r)]
    \,,
    \label{eq:vortexcondensate}
    \\
    A_{\theta} &= \frac{h(r)}{r} \, \textrm{diag}[-2a,a,a] \,,
    \label{eq:CFL_vortex_A}
\end{align}
\end{subequations}
with other gauge field components vanishing.
The radial functions $f(r)$, $g(r)$, and $h(r)$
all approach 1 as $r \to \infty$, and $f$ and $h$ vanish
at $r \eq 0$.
The gauge invariant order parameter
$\det(\phi/\Delta)$ approaches $e^{i\theta}$ far from the vortex core,
showing that this ansatz describes a vortex with minimal $U(1)_B$ winding,
or equivalently minimal superfluid circulation.
Such vortices are necessarily present when superfluid quark matter
rotates (above a critical frequency), as in neutron stars.
The energy per unit length of any straight, infinitely long superfluid vortex
is logarithmically IR divergent.
Minimizing the long distance energy density of the configuration
(\ref{eq:CFL_vortex}), proportional to
$
    r^{-2} [(1{-}2a)^2 + 2a^2]
$,
gives $a \eq 1/3$.%
\footnote
    {%
    Minimizing the complete energy determines the radial functions,
    which are monotonic and approach their asymptotic values
    exponentially rapidly on a scale set by the coherence length
    $\Delta^{-1}$ (for $f$ and $g$) or the much shorter $O((g \mu)^{-1})$
    color magnetic penetration length (for $h$)
    \cite{Balachandran:2005ev}.
    }

Now consider the $SU(3)$ holonomy $\Omega$ for some loop
$C$ in the presence of this minimal vortex running along the path $P$.
Assume that the curves $C$ and $P$ are everywhere widely separated
compared to the color magnetic penetration length.
Then by direct contour integration of the gauge field \eqref{eq:CFL_vortex_A}
one finds
\begin{align}
\label{eq:linking_phase_QCD}
    \gamma \equiv
    {\mathcal{N}}^{-1} \,
    \langle \tr \Omega[C]\rangle_{V[P]}
    = \exp [ \tfrac{2\pi i }{3} \> \ell(C,P)] \,,
\end{align}
where $\ell(C,P) \in \mathbb{Z}$ is the linking number of the
contours, and the normalization factor
$\mathcal{N} = \langle \tr \Omega[C]\rangle$
is the expectation value without the vortex.
This shows that
fundamental representation quarks and vortices have
$\mathbb{Z}_3$ braiding phases
in high density quark matter.

\section {Screening and fractionalization} 
How do the non-trivial 
color holonomies \eqref{eq:linking_phase_QCD} affect
physical quasiparticle excitations of dense QCD?
How does this relate to color screening
in a color superconductor?
These issues were considered in Ref.~\cite{Alford:2018mqj}
(see also Ref.~\cite{Chatterjee:2018nxe})
which examined CFL vortices and discussed possible implications for
quark-hadron continuity.
This followed the interesting earlier work in
Ref.~\cite{Cipriani:2012hr} 
which relied on the gauge variant
notion of color-magnetic flux to argue that
vortices could not continue smoothly between the CFL
and hadronic regimes.
The authors of
Ref.~\cite{Alford:2018mqj} correctly noted that gauge invariant
aspects of `color-magnetic flux' must be encoded
in color holonomies which encircle a vortex.
However, Ref.~\cite{Alford:2018mqj} asserted that such holonomies must be
trivial ``because the condensate, as a color triplet,
can completely screen the color charge of the probe quark."
This is inconsistent with the above explicit calculation
of the holonomy.
Moreover, while static test quarks are screened in QCD,
this only implies that the \emph{magnitude} of a large Wilson loop
will have perimeter-law behavior,
$|\langle \tr \Omega[C] \rangle | \sim e^{-\delta m \> |C|}$,
where $|C|$ denotes the perimeter and $\delta m$ is a
(renormalization scheme dependent) mass shift.
But such screening does not constrain the \emph{phase}
of the expectation which, as we have seen,
can be $\mathbb{Z}_3$-valued at high densities.

It is illuminating to recast this point in more physical terms
and examine how non-trivial holonomies affect
`screened' gauge invariant quasiparticle excitations.
The holonomy evaluated above represents the phase,
due to the color gauge field, acquired by a quark when encircling
a minimal vortex in the direction of the superfluid flow.
Such a quark,
whose minimal excitation energy is the pairing gap $\Delta$,
can be dressed by the diquark condensate $\phi$ to produce a
baryonic quasiparticle with the schematic structure
$q \phi$.
But the condensate is periodic, by construction, so
this cannot affect the holonomy induced $\mathbb Z_3$ phase 
experienced by the quark.    Equivalently, the condensate
    \eqref{eq:vortexcondensate} can be written as
    a product of $U(1)$ and $SU(3)$ factors;
    at large distance
    $\phi/\Delta \sim e^{i\theta/3} \times
    \mathrm{diag}[e^{2i\theta/3},e^{-i\theta/3},e^{-i\theta/3}]$.
    The angular variation of the $SU(3)$ factor cancels
    the color holonomy of the quark --- but the $U(1)$
    factor reinstates exactly the same overall phase change.

This $\mathbb Z_3$ phase implies that a $q\phi$ excitation
has anyonic statistics with superfluid vortices.
Just as in BCS superconductors \cite{Wilczek:1981du},
one can also interpret the holonomy induced phase
$\alpha \equiv 2\pi/3$ as a shift in the allowed values
of (orbital) angular momentum of this excitation
in the presence of a vortex,
$\alpha = 2\pi \Delta L_z/\hbar$,
so $\Delta L_z = \hbar/3$.
More precisely, for excitations far from the vortex core
    it is the azimuthal component of the
    ``kinetic'' or ``covariant'' angular momentum,
    $
	L_z \equiv \hat z \cdot \vec r \times (\vec p - \vec A)
    $,
    which is fractionalized.
    This differs from the conserved angular momentum,
    or generator of rotations,
    which remains integer quantized.
    However, it is the kinetic angular momentum $L_z$ which appears in the
    relation between angular velocity and angular momentum,
    $
	d\theta/dt = L_z/(m r_\perp^2)
    $,
    or equivalently in the rotational component of kinetic energy,
    $
	L_z^2 / (2 m r_\perp^2)
    $.
    In other words, a $q\phi$ excitation far from the vortex
    moves as if it is a free particle
    (in the absence of any gauge holonomy)
    with fractional angular momentum, $L_z/\hbar \in \mathbb Z + 1/3$.

\begin{table}
  \begin{tabular}{c|c|c|c}
    $\Delta L_z$ & $0$ & ~~$+\hbar/3$~~ & ~~$-\hbar/3$~~
    \\ \hline
    \textrm{bosons} & $q\bar{q}$ & $\bar{q}\bar{q} \phi$ & $qq \phi^{*}$
    \\[2pt] \hline
    \textrm{fermions} & ~$qqq$,\, $\bar{q}\bar{q}\bar{q}$~ & $q \phi$ & $\bar{q} \phi^{*}$ \\
  \end{tabular}
  \caption{%
  Shifts in orbital angular momentum, $\Delta L_z$, of 
  low energy excitations in the CFL phase in the presence of a minimal
  superfluid vortex.
  Here $q$ denotes a quark excited above the Fermi surface,
  $\bar{q}$ denotes a corresponding hole,
  and $\phi$ is the diquark condensate.
  Each indicated combination can form a physical,
  gauge invariant excitation.
  }
  \label{table:phases}
\end{table}

Table~\ref{table:phases}
lists the analogous shifts in $L_z$ for other possible 
quasiparticles in CFL color superconductors.
Note that these considerations imply the existence
of sharply distinct classes of both baryonic and mesonic
excitations in the presence of a vortex.

\section {Topology in effective theory}
We obtained the $\mathbb Z_3$ braiding phases \eqref{eq:linking_phase_QCD}
using the microscopic degrees of freedom of the QCD Lagrangian.
However, the topological data in these braiding phases
must also be encoded in any correct long distance 
description of the system.
Since these braiding phases are insensitive to a bare quark mass
which gaps out other NGBs (pions) associated with chiral
symmetry breaking,
this data must appear in the minimal effective field theory (EFT)
describing the dynamics of $U(1)_B$ Nambu-Goldstone bosons
and the response of various probes to superfluid fluctuations.
The conventional effective action \eqref{eq:naiveEFT} cannot
reproduce these braiding phases, and thus cannot be complete.
What must be added to the EFT to fix this problem?

The correct EFT must reproduce the holonomy-vortex
linking relation (\ref{eq:linking_phase_QCD}).
When considering the theory on a non-simply connected space
(such as a toroidal compactification),
it should also reproduce correct $\mathbb Z_3$-valued holonomies
when there is superfluid flow around a non-contractible cycle.
To construct such a theory, a natural starting point is a
4D topological quantum field theory (TQFT) known as BF theory
\cite{Horowitz:1989ng,Blau:1989bq,Banks:2010zn,Kapustin:2014gua}.
This TQFT has the Euclidean action
\begin{align}
    S_{\textrm{BF}} = \frac{i p}{2\pi} \int_{M_4} b_2 \wedge d a \,,
\end{align}
where $M_4$ is a four-dimensional spacetime manifold,
$b_2$ is a $2$-form gauge field, \footnote{A $p$-form field is a totally antisymmetric rank-$p$ tensor, and may be integrated over any dimension $p$ manifold without needing a metric. }
$a$ is a $1$-form gauge field,
$d$ denotes exterior differentiation,
and $p \in \mathbb{Z}$.
By itself, this action describes a $\mathbb Z_p$ discrete gauge theory
\cite{,Banks:2010zn,Kapustin:2014gua}.
We wish to relate holonomies of the gauge field $a$ to the superfluid 
circulation.
For holonomies encircling vortices, this may be accomplished by
adding the term
$
    \frac{1}{2\pi i}\int_{M_4} b_2 \wedge d^2 \varphi
$
to $S_{\rm BF}$.
The two-form $d^2\varphi$ is proportional to the superfluid vorticity
and vanishes everywhere except at a vortex center where
$\varphi$ is ill-defined.
The resulting equation of motion,
$
    p \, da = d^2\varphi
$,
connects the Abelian field strength $da$ to the vorticity.
By Stokes theorem, this is the same as equating the
circulation $\oint_C d\varphi$ with $p$ times the
holonomy phase $\oint_C a$.
For $p\eq 3$, this is precisely our linking number
relation \eqref{eq:linking_phase_QCD}.

These considerations suggest that a proper low energy effective
action for high-density QCD should be
\begin{align}
    \!\!
    S_{\rm eff}
    \mathop{=}^?
    \int_{M_4}
    \tfrac{f^2}{2} \, d \varphi \wedge \star d\varphi
    +  \tfrac{i}{2\pi}
    \int_{M_4}
    b_2 \wedge ( p \, d a -  d^2 \varphi) \,,
\label{eq:Seff1}
\end{align}   
where $\star$ denotes the Hodge dual and
we chose units where $v_s \eq 1$.
But this effective theory is not fully correct,
as it cannot reproduce the correlation
between holonomies and superfluid flow when spacetime
is not simply connected,
as in our earlier example of $\mathbb{R}^3 \times S^1$.
We need an effective action that can connect integrals
of $a$ and $d\varphi$ around non-contractible curves.
However, this is impossible with a
local gauge-invariant 4D effective action.

This problem has a natural solution inspired by the Wess-Zumino-Witten
term of chiral perturbation theory \cite{Wess:1971yu,Witten:1983tw}.
Let $M_{5} = M_4 \times [0,1]$ denote the 5D manifold which is the product
of $M_4$ times an interval,
and let $w \in [0,1]$ be a coordinate along this interval.
Regard the $w\eq 1$ boundary as the physical 4D spacetime
and extend the superfluid condensate phase $\varphi$ 
to a function $\widetilde\varphi$ on $M_5$
(i.e., a mapping of $M_5$ into $S^1$)
which coincides with $\varphi$ at $w \eq 1$,
is constant at $w \eq 0$,
and is smooth and differentiable almost everywhere in the interior of $M_5$.
When the phase $\varphi$ winds around a physical vortex in $M_4$,
the vortex worldsheet in $M_4$ will extend to a 3D worldvolume in $M_5$
on which $\widetilde\varphi$ is ill-defined
(and around which $\widetilde\varphi$ has non-zero winding).
If $\varphi$ has winding around some non-contractible cycle
in $M_4$, then $\widetilde\varphi$ will necessarily be ill-defined on some
3D ``vortex worldvolume'' in the interior of $M_5$.
(If $M_4$ is closed, this 3-surface is also closed.)
We now replace the incomplete effective theory (\ref{eq:Seff1})
with a 5D BF theory coupled to the vorticity of $\widetilde\varphi$,
\begin{align}
    \!\!
    S_{\rm eff} =
    \int_{M_4} \tfrac{f^2}{2} \, d\varphi \wedge \star d\varphi
    + \tfrac{i}{2\pi} \int_{M_5} \! b_3 \wedge
	(p \, d a - d^2\widetilde\varphi ) \,.
\label{eq:QCD_TQFT}
\end{align}
Here $b_3$ is a 3-form gauge field which is required to vanish
at $w \eq 1$, while the gauge field $a$ is required to vanish at $w \eq 0$.%
\footnote
    {%
    The action \eqref{eq:QCD_TQFT} shifts by integer multiples of
    $2\pi i p$ under the gauge transformations
    $b \to b + d \lambda_{(2)}$ and
    $a\to a + d\lambda_{(0)}$, where $\lambda_{(2)}$ and $\lambda_{(0)}$
    are 2-form and 0-form gauge functions constrained to vanish at
    $w=1$ and at $w=0$ respectively.  Gauge invariance of the functional
    integral then implies that $p \in \mathbb{Z}$.
    }
The 5D equations of motion require that
$p \, da \eq d^2\widetilde\varphi$ and this, once again,
implies that the circulation $\oint_C d\varphi$ around any
closed curve $C$ in $M_5$ coincide
with $p$ times the holonomy phase $\oint_C a$.
And, importantly, this now includes non-contractible curves lying on
the boundary at $w \eq 1$.

Consequently, the effective theory \eqref{eq:QCD_TQFT}, involving
a 5D topological term coupled to a 4D superfluid effective action, 
correctly reproduces the association between the $\mathbb Z_3$
holonomy and superfluid circulation in all geometries.   The value of writing down 
Eq.~\eqref{eq:QCD_TQFT} is simply to establish that it is possible to construct 
an effective action describing the dynamical $U(1)_B$ Goldstone boson field $\varphi$ 
and some emergent color-singlet fields $b$ and $a$, whose own dynamics are 
essentially trivial.  The role of these emergent fields, which enter the new 5D term is simply to encode the non-trivial gauge-invariant 
holonomies of the color gauge fields which we found using a more
 microscopic description of the theory.

\section{Implications}
%
If the quark-hadron continuity conjecture of Ref.~\cite{Schafer:1998ef}
is correct, 
then $\mathbb Z_3$ particle-vortex braiding phases and
associated angular momentum fractionalization must persist
as the density decreases all the way down to the onset of
superfluidity in nuclear matter,
at least sufficiently close to the $SU(3)$ flavor symmetric limit.  
This would imply that hadronic nuclear matter \emph{cannot} be accurately described by a local effective theory
involving colorless baryon and meson degrees of freedom, which is very hard to believe.   Such a result would not be consistent with the standard picture that in low density nuclear matter, test quarks are screened by pair production of quark-antiquark pairs from the vacuum.    The baryon-number violating condensate in nuclear matter cannot screen a test color charge, because this condensate has the quantum numbers of a dibaryon and is color-neutral.  This should be contrasted with the situation in dense quark matter, where a diquark condensate can screen test quarks, leading to the explicit results described above concerning color holonomies that encircle superfluid vortices and associated fractionalization of orbital angular momentum.

The most plausible interpretation of our results is that quark-hadron continuity fails,
with at least one phase transition separating
high density quark matter, with its $\mathbb Z_3$ topological features,
from lower density superfluid nuclear matter lacking these features.
To understand why a discrete change in particle-vortex braiding statistics
    implies a thermodynamic phase transition,
    note that the ground state must have non-zero amplitudes
    for configurations in which vortex loops are present.
    Any such loop affects the quasiparticle spectrum in its vicinity,
    and hence affects fluctuation corrections of the ground state energy.
    Consequently, an abrupt change in particle-vortex braiding statistics
    should generate non-analyticity in the ground state energy.
    
We emphasize that such a phase transition in dense nuclear
matter does not conflict with the well-known Fradkin-Shenker results
on Higgs-confinement complementarity \cite{Fradkin:1978dv},
which demonstrate continuity between confinement and Higgs regimes in
systems \emph{without} a $U(1)$ global symmetry.  In contrast, the
presence of spontaneously broken $U(1)_B$ symmetry and the consequent
existence of superfluid vortex excitations plays a central role in our
discussion of particle-vortex statistics.

Finally, we emphasize
we have assumed coinciding quark masses and
$SU(3)_V$ flavor symmetry throughout this paper.
This flavor-symmetric limit was the setting for the quark-hardron continuity conjecture of Ref.~\cite{Schafer:1998ef}.
Of course, the real world has distinct quark masses,
plus electromagnetic and weak interactions.
For asymptotically large $\mu$, such flavor-breaking effects
are negligible \cite{Alford:2007xm}, but as $\mu$ decreases
flavor symmetry breaking effects grow in importance.
A detailed analysis of flavor-breaking effects is undoubtedly interesting,
but is left to future work.  Aside from the practical consideration of allowing a discussion of the physical point, an  analysis of flavor symmetry breaking perturbations  is also important to address theoretical points of principle.  For  example, as it stands, it is tempting to \emph{define} the density-driven phase transition between nuclear matter and quark matter discussed above as a color confinement-deconfinement transition.  But to make such a definition reasonable one would want to be sure that that the transition persists for a range of flavor-symmetry breaking perturbations. 

\section {Outlook}

The recognition that high density quark matter exhibits
non-trivial topological features such as 
$\mathbb Z_3$ particle-vortex braiding statistics,
angular momentum fractionalization,
and an emergent $\mathbb Z_3$ discrete gauge field,
shows that color superconductivity has novel physical
signatures which are distinct from the more general
phenomenon of ordinary superfluidity.
This realization leads to numerous additional questions.
What are observable signatures of angular momentum fractionalization
around vortices?
How do $\mathbb Z_3$ particle-vortex statistics affect 
quasiparticle dynamics and transport processes in rotating quark matter?
If quark-hadron continuity fails, what is the nature of the
phase transition(s) which separate these regimes?
How do superfluid vortices in a rotating neutron star
behave at such a phase interface \cite{Cipriani:2012hr}?
Are there superfluids in condensed matter systems with
similar particle-vortex statistics?
We hope future work can shed light on these and related questions.

{\bf Acknowledgments.}
We thank Mark Alford, Paulo Bedaque, Tyler Ellison, Zohar Komargodski, Sanjay Reddy, Martin Savage,
Sujeet Shukla, Boris Spivak, Naoki Yamamoto, and especially Lukasz Fidkowski for
helpful conversations.
This work was supported, in part,
by the U.~S.~Department of Energy grants 
DE-SC\-0011637 and DE-FG02-00ER-41132.

{\bf Note added.}
After our work was posted on the arXiv, the paper  \cite{Hirono:2018fjr} appeared, arguing that quark-hadron continuity is consistent with an unbroken "emergent two-form symmetry."  We disagree with a number of assertions in this paper and its conclusions.  However, nothing in Ref.  \cite{Hirono:2018fjr} casts any doubt on, or even addresses, our explicit evaluation of non-trivial $\mathbb{Z}_3$ holonomies associated with superfluid circulation in high density QCD and the resulting physical effects discussed above.

\bibliography{small_circle}
\end{document}